\documentclass[runningheads]{svmult}

\usepackage{makeidx}   
\usepackage{graphicx}  
\usepackage{multicol}  
\usepackage{physprbb}  
\makeindex             

\begin{document}
\title*{ The effect of the Galactic gas distribution on the expected
Cosmic Rays spectrum}
\toctitle{The effect of the Galactic gas distribution
\protect\newline on the expected Cosmic Rays spectrum}
%
%
\titlerunning{The effect of the gas distribution on the CR spectrum}
%
\author{Mercedes Moll\'{a}
\and Manuel Aguilar
\and Juan Alcaraz
\and Javier Berdugo
\and Jorge Casaus
\and Carlos D\'{\i}az
\and Elisa Lanciotti
\and Carlos Ma\~{n}\'{a}
\and Jes\'{u}s Mar\'{\i}n
\and Gustavo Mart\'{\i}nez
\and Carmen Palomares
\and Eusebio S\'{a}nchez
\and Ignacio Sevilla
\and Ana Sof\'{\i}a Torrent\'{o}}
\authorrunning{Moll\'{a} et al.}
%
%
\institute{CIEMAT, Avda. Complutense 22, 28040 Madrid (Spain)}

\maketitle              

\section{Introduction}

\begin{figure}[b]
\begin{center}
\includegraphics[width=0.4\textwidth,angle=-90]{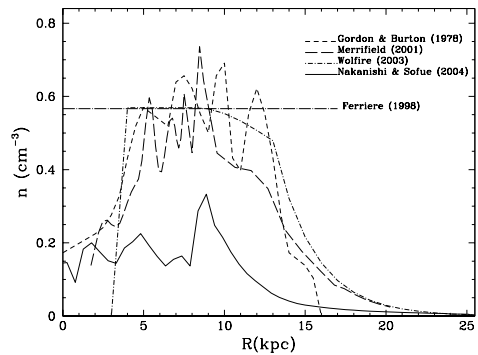}
\end{center}
\caption[]{The radial distribution of the diffuse gas atomic density
as obtained from \cite{joun6} compared
with other author's distributions as labeled in the figure}
\label{gas}
\end{figure}

Cosmic ray (CR) nuclei are accelerated particles which move randomly
through the interstellar medium (ISM), where they suffer scattering,
reacceleration and energy loss processes before reaching
Earth. Spallation processes also take place forming secondary nuclei
by fragmentation of heavier ones. Due to the impossibility of
observing directly their original direction, the determination of
possible sources where these particles originated requires the use of
codes to simulate the propagation of CR within the Galaxy. This
consists of a spiral disk with a thickness of 2h $\sim$ 200 pc, where
CR are created, and a halo with a height H, where they diffuse.

From the existing data, mostly $\rm ^{10}Be/^{9}Be$ and $\rm B/C$, it
is deduced that CR go through a mean density of $\sim 0.3$ cm$^{-3}$
(\cite{joun4,joun5}).  Since the total density of
the ISM is considered as 1 $\rm cm^{-3}$, ($\rm <n_{HI}> \sim 0.5 \,
cm^{-3}$; \cite{joun1,joun2,joun7,joun10}), 
a large effective halo and/or a local
bubble or cavity of low density are required. This result is very
dependent on the actual disk ISM density, largely uncertain. 
\cite{joun6} have very recently obtained a new map of the diffuse
gas distribution (Fig.~\ref{gas}), from which $\rm <n_{HI}> \sim 0.2
\, cm^{-3}$, a factor of 2 lower than previous estimates.

\section{Computed Models and Results} 

\begin{figure}[b]
\centerline{%
\begin{tabular}{c@{\hspace{2pc}}c}
\includegraphics[width=0.4\textwidth,angle=0]{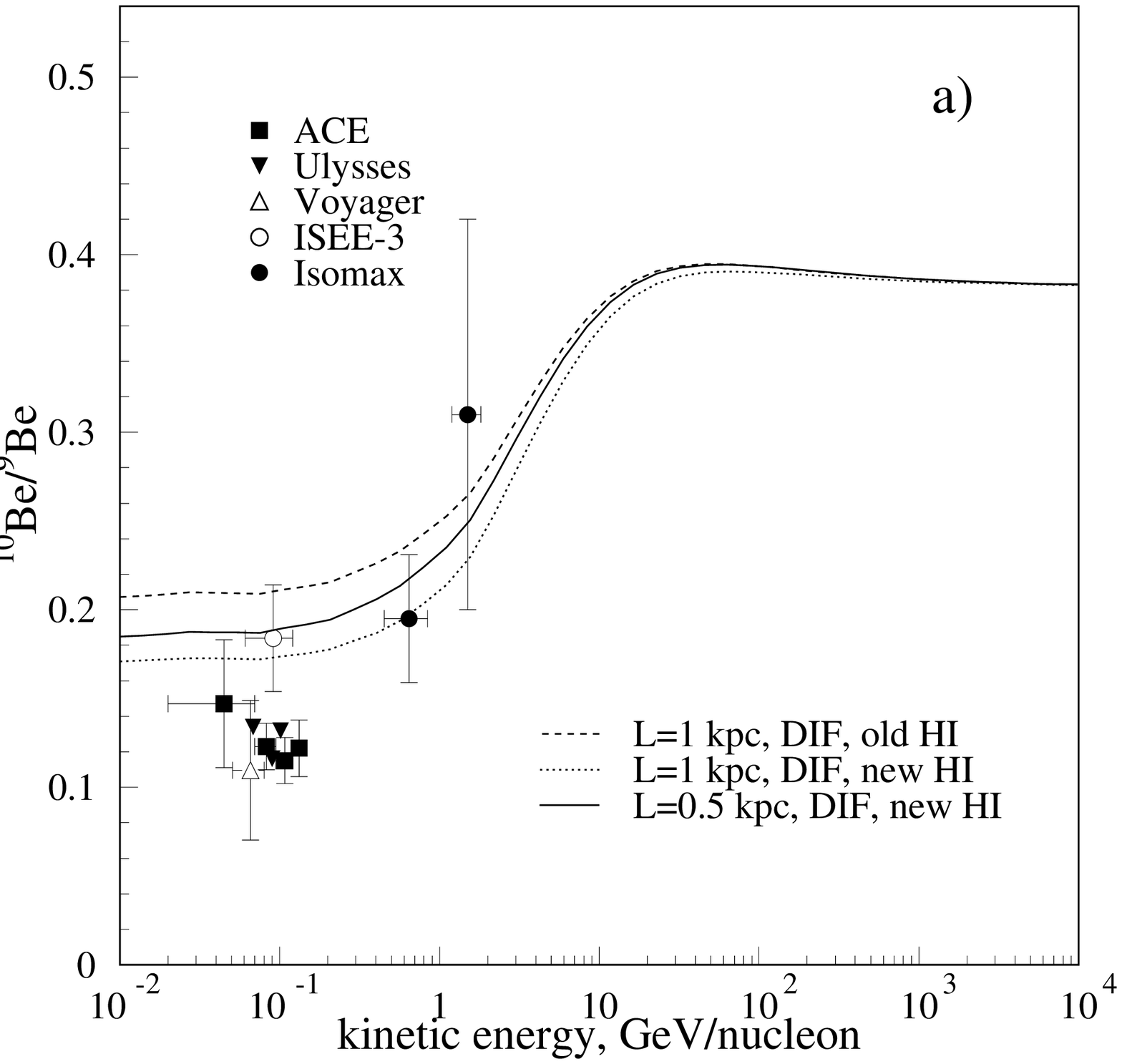} &
\includegraphics[width=0.4\textwidth,angle=0]{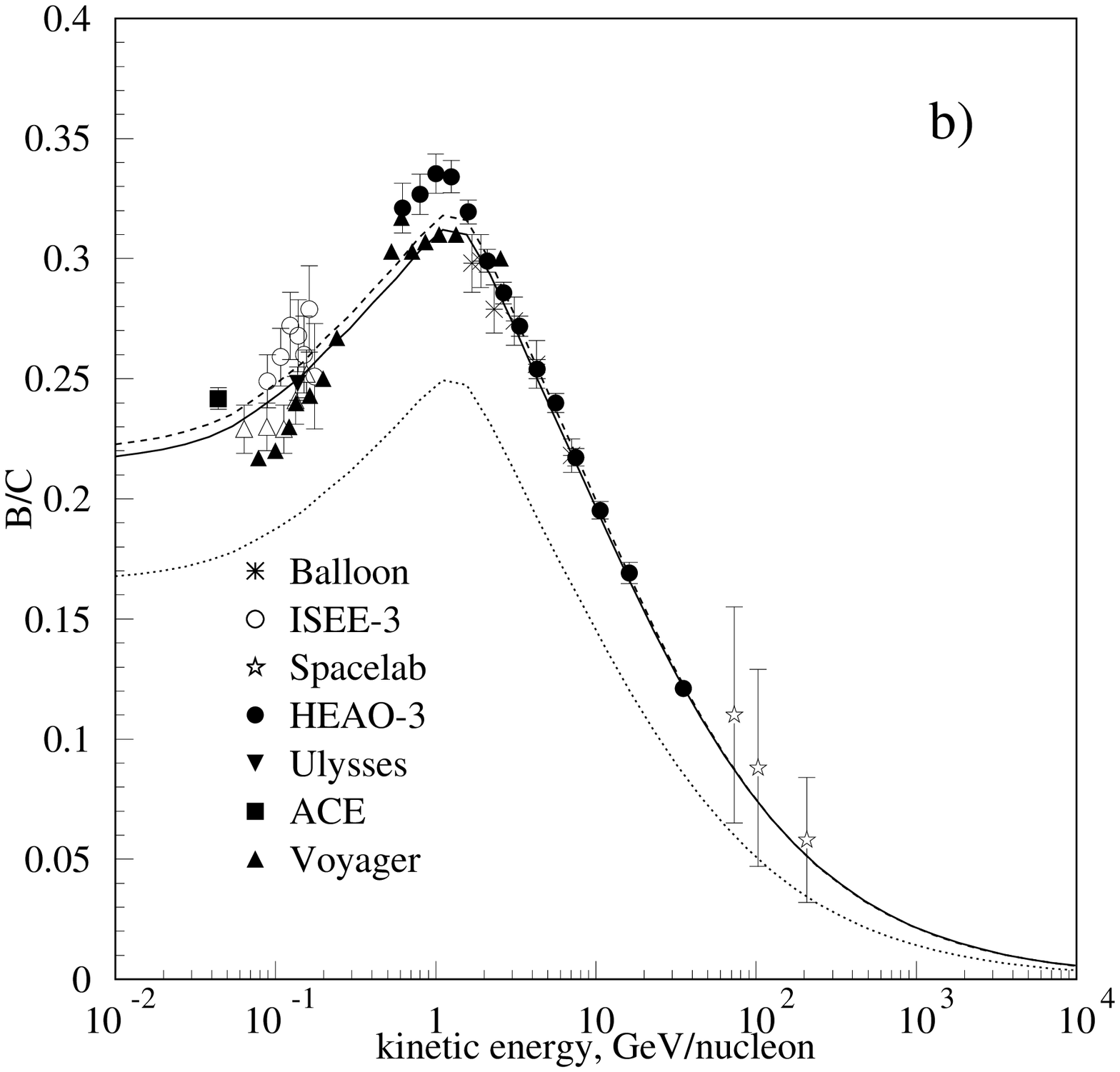}\\
\end{tabular}}
\caption[]{The spectra of a) $\rm ^{10}Be/^{9}Be$, and b) of $\rm
B/C$.  Models are represented by lines as labeled}
\label{bebc}
\end{figure}
We now use the GALPROP code, developed by \cite{joun8,proc1}, 
to compute how CR propagates.  The model includes a
realistic diffuse gas radial distribution (\cite{joun3}).
A halo height $H=1-5$ kpc is obtained, depending of the model.  
Results for $\rm ^{10}Be/^{9}Be$ and $\rm B/C$ are shown with 
dashed lines in Fig.~\ref{bebc} a) and b), respectively, for 
a diffusion model with $H=1$ kpc.

We now change the old HI distribution with the new one.  The dotted
lines in Fig.2 represent the same model using this new distribution.
It is evident that it is not yet valid.  The solid lines correspond to
a new good enough model, with a halo height $\rm H=0.5$ kpc. This
implies that CR diffuse mostly in the disk closer to the sources than
it was deduced before.  This result agrees with \cite{joun9}
who obtain distances as short as 500 pc between the source and the
Sun, from the ACE radionuclide measurements.

%


\begin{thebibliography}{8.}
\addcontentsline{toc}{section}{References}

\bibitem{joun1}Dickey, J. M. \& Lockman, F.J. 1990, A.R.A.A.,
\textbf{28}, 215
\bibitem{joun2}Ferri\'{e}rre, K. 1998, ApJ., \textbf{497}, 759
\bibitem{joun3}Gordon, M.A.  \& Burton, W.B. 1976, ApJ., \textbf{208},
346
\bibitem{joun4} Jones, F.~C., Lukasiak, A., Ptuskin, V., \& Webber, W.
\ 2001, ApJ, \textbf{547}, 264
\bibitem{joun5}Maurin, D., Donato, F., Taillet, R. \& Salati, P. 2001,
ApJ., \textbf{555}, 585
\bibitem{joun6}Nakanishi, H. \& Sofue, Y.  2004, PASJ, \textbf{55},
191
\bibitem{joun7}Olling, R.~P.~\& Merrifield, M.~R.\ 2001, MNRAS, 
\textbf{326}, 164
\bibitem{joun8}Strong \& Moskalenko 1998, ApJ, \textbf{509}, 212
\bibitem{proc1}Strong \& Moskalenko 2003, 28 IRC Conf., 1921
\bibitem{joun9}Yanasak, N.E., Wiedenbeck, M. E., Mewaldt, R.A. et
al. 2001, ApJ., \textbf{563}, 768 
\bibitem{joun10}Wolfire, M.~G., McKee, C.~F., Hollenbach, D., et al.\
2003, ApJ, \textbf{587}, 27

\end{thebibliography}
\end{document}